# Equilibration of a polycation – anionic surfactant mixture at the water/vapor interface

Andrew Akanno,[1,2] Eduardo Guzmán,[1,2,*] Laura Fernández-Peña,[1] Sara Llamas,[1] Francisco Ortega,[1,2] and Ramón G. Rubio[1,2,*]

[1] Departamento de Química Física-Universidad Complutense de Madrid, Ciudad Universitaria s/n, 28040 Madrid, Spain

[2] Instituto Pluridisciplinar-Universidad Complutense de Madrid, Paseo Juan XXIII, 1, 28040 Madrid, Spain

*To whom correspondence should be sent: rgrubio@quim.ucm.es (RGR) and eduardogs@quim.ucm.es (EG).

PDADMAC – SLES mixtures

**Abstract**

The adsorption of concentrated poly(diallyldimethylammonium chloride) (PDADMAC) - sodium lauryl ether sulfate (SLES) mixtures at the water / vapor interface has been studied by different surface tension techniques and dilational visco-elasticity measurements. This work tries to shed light on the way in which the formation of polyelectrolyte – surfactant complexes in the bulk affects to the interfacial properties of mixtures formed by a polycation and an oppositely charged surfactant. The results are discussed in terms of a two-step adsorption-equilibration of PDADMAC – SLES complexes at the interface, with the initial stages involving the diffusion of kinetically trapped aggregates formed in the bulk to the interface followed by the dissociation and spreading of such aggregates at the interface. This latter process becomes the main contribution to the surface tension decrease. This work helps on the understanding of the most fundamental bases of the physico-chemical behavior of concentrated polyelectrolyte – surfactant mixtures which present complex bulk and interfacial interactions with interest in both basic and applied sciences.

# 1. Introduction

The broad range of technological and industrial applications involving polymer – surfactant mixtures justifies the extensive research performed for unravelling the physico – chemical bases underlying the bulk behavior and adsorption at interfaces, both fluid and solid, of this type of systems.[1-8] In recent years, special interest has been paid to the study of mixtures involving polyelectrolytes and oppositely charged surfactants which has been used for long time in cosmetics, food science and drug delivery.[7, 9]

Despite the large number of papers, several aspects of their behavior remain controversial, mainly due to the fact that the polyelectrolyte – surfactant complexes formed are non-equilibrium aggregates.[4, 10-14] This plays an important role on the adsorption of such mixtures to fluid interfaces. The adsorption of polyelectrolyte – surfactant mixtures has been traditionally studied using surface tension measurements.[15-18] However, it was not until the development of neutron reflectometry, when it was possible to obtain a deeper understanding of the behavior of polyelectrolyte – surfactant interfacial layers.[19] Penfold's group was the first to report the correlations existing between the adsorption of polyelectrolyte – surfactant mixtures at fluid interfaces and the aggregation phenomena existing in the bulk.[20-23] However, these studies neglected the existence of non-equilibrium effects, providing a description of the adsorption of complexes at the fluid interface in terms of an extended Gibbs equation of state which was able to account by the non-regular dependences of the surface tension on the concentration (surface tension peaks).[24-25] This issue was solved by Campbell et al.[13, 26-30] who provided a comprehensive description of the behavior of interfacial layers formed by polyelectrolyte – surfactant mixtures, including a detailed discussion of the importance of the non-equilibrium effects and the bulk phase – behavior in the adsorption of polyelectrolyte – surfactant mixtures at fluid interfaces.[10] However, the PDADMAC – SLES mixtures complete description of the physico-chemical behavior of polyelectrolyte – surfactant mixtures at fluid interfaces requires also a careful examination of the response of such systems to mechanical perturbations.[31-36] This is because most technological and industrial problems involving polyelectrolyte – surfactant mixtures, e.g. foams stabilization,[37] are related to their response to dilation or shear deformations. The first studies dealing with the rheological properties of this type of systems were carried out by Regismond et al.[38-39] who found a strong synergetic effect of the polyelectrolyte – surfactant interaction in the interfacial properties. Later, Bhattacharyya et al.[17] and Monteux et al.[40] analyzed the rheological response of polyelectrolyte – surfactant mixtures at fluid interfaces to obtain correlation with their capacity on the stabilization of foams. They found that the adsorption of polyelectrolyte-surfactant mixtures to the fluid interface leads to the formation of layers which present a gel-like character. The formation of such type of layers hinders bubble coalescence and foam drainage. More recently, Noskov et al.[13-14, 41-44] studied extensively the response against dilation of interfaces formed by several mixtures of polyelectrolytes and surfactants, concluding that the dilational response of these systems was mainly related to the heterogeneity of the adsorbed layers. Such heterogeneity depends on the structure of the complexes formed in the bulk.

Most studies in the literature are devoted of the study of mixtures containing low polymer concentrations. However, most consumer products contain complex mixtures in which polymer concentration overcomes the overlapping one, $c^*$,[45-46] which affect the physico – chemical properties of the system,[47] as it was demonstrated in our previous studies.[48-50] The main difference between concentrated and diluted mixtures is associated with the different degree of binding of surfactant molecules to polymer chains. For concentrated solutions are considered a high binding of surfactant molecules to polymer chains is found, with the

PDADMAC – SLES mixtures

remaining amount of surfactant free in solution being rather small (below 10%), even for those mixtures with surfactant concentrations close to the phase separation region. Thus, it is possible to assume that concentrated mixtures contain only complexes, and the role of the free surfactant in their behavior can be considered negligible. On the other side, the binding degree is lower in diluted mixtures, thus both polyelectrolyte – surfactant complexes and free surfactant molecules are present in solution. The different bulk aggregation found for concentrated and dilutes mixtures (c << c*) is reflected on their adsorption to fluid interfaces, especially in the interfacial composition of the adsorbed layers. Concentrated mixtures lead upon adsorption to the formation of interfacial layers mirroring the composition of the bulk solutions. However, this is not true when diluted mixtures are considered. [10, 49]

This work deals with the study of the equilibrium and dynamics properties of interfacial layers formed upon adsorption of poly(diallyldimethylammonium chloride) (PDADMAC) and sodium lauryl ether sulfate (SLES) to the water / vapor interface. This mixture has been chosen because SLES is a surfactant that is introduced in cosmetic products for replacing the sodium dodecyl sulfate (SDS), which has been the traditionally studied surfactant. This is due to the less irritant character and lower susceptibility to degradation of SLES. [9, 51] For the aim of this study, surface tension has been measured using different tensiometers. Surface tension measurements can be considered a powerful tool to study the non-equilibrium effects appearing during the adsorption of polymer – surfactant mixtures at fluid interfaces. [49] Furthermore, the adsorption kinetics of the complexes at the water / vapor interface has been followed by the evolution of the dynamic surface tension evaluated using a drop shape tensiometer, whereas the mechanical response of the interfaces to dilational perturbations has been studied by oscillatory barrier experiments in a Langmuir trough.[52] The study of the interfacial properties and the bulk aggregation phenomena is expected to provide important

PDADMAC – SLES mixtures

insights for detangling the complex scenario associated with the formation of interfacial layers of polyelectrolyte - surfactant mixtures.

## 2. Experimental Section

**2.1 Chemicals.** PDADMAC with an average molecular weight in the 100 – 200 kDa range was purchased from Sigma-Aldrich (Germany), and was used without further purification. SLES was purchased from Kao Chemical Europe S.L. (Spain) and was purified by recrystallization from ethanol.[53] Ultrapure deionized water used for cleaning and solution preparation had a resistivity higher than 18 MΩ·cm and a total organic content lower than 6 ppm (Younglin 370 Series, South Korea).

The pH of all solutions was adjusted to 5.6 using glacial acetic acid and the ionic strength was kept constant by adding 0.3 wt% of KCl (purity > 99.9 %). The PDADMAC concentration was kept constant at 0.5 wt% in all the samples. The pH, ionic strength and polymer concentration has been chose to mimic the conditions appearing in most cosmetic formulations for hair care applications in which are involved the here studied mixtures, among other components.[9]

**2.2. Samples preparation**

Polyelectrolyte–surfactant mixtures were prepared following the procedure described by Llamas et al.[49] which can be summarised as follows: first the required amount of PDADMAC aqueous stock solution (concentration 20 wt%) for preparing final solutions with polymer concentration around 0.5 wt% is weighted and poured into a flask. Afterwards KCl is added to reach a total salt concentration of 0.3 wt% in the final sample.

PDADMAC – SLES mixtures

In the last step, the addition of the SLES solution (pH ~ 5.6), and final dilution with acetic acid solution of pH ~ 5.6 to reach the desired bulk compositions is carried out. In all the cases, the SLES solutions used in the preparation of polyelectrolyte – surfactant mixtures presents a surfactant concentration one order of magnitude higher than the final sample. Once the samples were prepared, a mild stirring of the samples (1000 rpm) using a magnetic stirrer during one hour was carried out to ensure the samples homogeneization. This procedure is expected to lead to the formation of kinetically-trapped aggregates during the mixing process due to the establishment of local concentration gradients,[26] thus to ensure reproducibility in the results all the samples were aged during one week before their use.

**2.3. Techniques**

*a. Surface tension measurements*

The surface tension dependence on the surfactant concentration was measured for SLES and PDADMAC – SLES solutions using different tensiometers. The adsorption at the water / vapor interface was measured until steady state was reached, i.e. changes of surface tension smaller than 0.1 mN·m$^{-1}$ during 30 minutes. Special care was taken to minimize evaporation effects. Each reported experimental data was the average of at least three independent measurements. All experiments were carried out at 25.0 ± 0.1°C. Further experimental details can be found in our previous publication.[49]

*a.1. Surface force tensiometers*. A surface force tensiometer from Krüss K10 (Germany) was used with a Pt Wilhelmy plate as contact probe. Additionally, a surface force tensiometer from Nima Technology (United Kingdom) was used with disposable paper plates (Whatman CHR1 chromatography paper).

PDADMAC – SLES mixtures

*a.2. Shape profile analysis tensiometer.* A home-made profile analysis tensiometer in pendant drop configuration was also used to measure the surface tension at the water / vapor interface.[54] This tensiometer also allows us to follow the time evolution of the surface tension providing information about the adsorption kinetics at the fluid interface.

*b. Dilational rheology*

A Langmuir balance NIMA 702 from Nima Technology (United Kingdom) equipped with a surface force tensiometer was used for measuring the surface tension response to sinusoidal changes on the surface area performed following the oscillatory barrier method.[52, 55] Following this methodology was possible to obtain the dilational viscoelastic moduli $\varepsilon^* = \varepsilon' + i\varepsilon''$ ($\varepsilon'$ is the dilational elastic modulus and $\varepsilon''$ is the viscous modulus) in the frequency range $10^{-1} - 10^{-2}$ Hz and at area deformation amplitude $\Delta u = 0.1$, which was checked to be an appropriate value to ensure results within the linear regime of the layer response.

*c. Zeta potential measurements*

A Zetasizer Nano ZS instrument (Malvern Instruments Ltd., Worcestshire, UK) was used for the measurement of the electrophoretic mobility, $\mu_e$.

*d. Dynamic Light Scattering*

The Zetasizer Nano ZS was also used for dynamic light scattering (DLS) measurements. Using DLS it is possible to estimate the *apparent* hydrodynamic radius, $R_H^{app}$, following the approach described in our previous work.[48]

*d. Turbidimetry*

PDADMAC – SLES mixtures

The turbidity of the PDADMAC – SLES mixtures was determined measuring the transmittance of the mixtures at 400 nm using a UV/visible spectrophotometer (HP-UV 8452). The turbidity is represented as [100-T (%)]/100, with T being the transmittance.

*e. Binding isotherm*

The binding isotherm of SLES to PDADMAC was determined by potentiometric titration using a surfactant selective electrode model 6.0507.120 from Metrohm (Switzerland). The binding degree of surfactant $\beta$ is estimated from the potentiometric measurements as[56]

$$\beta = \frac{c_s^{free}}{c_{monomer}} \qquad (1)$$

with $c_s^{free}$ being the concentration of free surfactant and $c_{monomer}$ the concentration of charged monomers of the polyelectrolyte.

## 3. Results and discussion

### 3.1. PDADMAC – SLES interaction in solution

The first step on the study of PDADMAC – SLES involves understanding the interactions occurring in the bulk between polyelectrolyte segments and surfactant molecules. The study of the aggregation phenomena occurring in bulk provides important information for a better understanding the adsorption of polyelectrolyte – surfactant mixtures at fluid interfaces. Figure 1a shows the dependence of the electrophoretic mobility, $\mu_e$, on the bulk surfactant concentration for PDADMAC – SLES mixtures.

PDADMAC – SLES mixtures

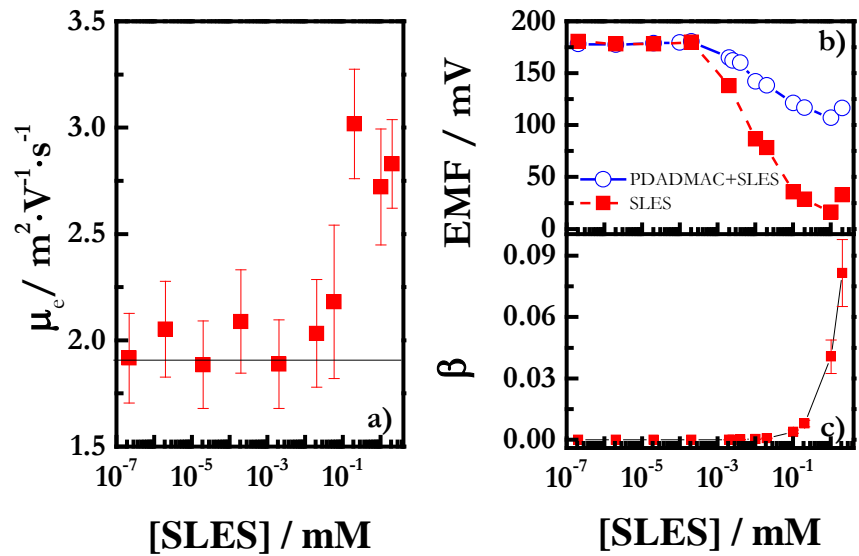

Figure 1. (a) SLES concentration dependence of the $\mu_e$ for PDADMAC – SLMT mixtures. The horizontal solid line represents the $\mu_e$ value of PDADMAC solutions (0.5 wt%). (b) Electromotive force, EMF, dependence on the SLES concentration for PDADMAC – SLES mixtures and for pure SLES solutions. (c) Binding isotherm for PDADMAC – SLES mixtures as was obtained from the analysis of the results in panel b.

The $\mu_e$ of PDADMAC – SLES mixtures shows positive values, close to that corresponding to pure polyelectrolyte with the same concentration, within the entire range of explored bulk surfactant concentrations. This can be understood considering that the number of surfactant molecules is not enough to neutralize the charge of the polyelectrolyte monomers, thus the complexes formed in the bulk are undercompensated, containing an excess of positively charged monomers. A simple calculation allows one to estimate that the number of available monomers for surfactant molecule is around 36 for SLES concentration about 1 mM, i.e. there are 35 positively charged monomers for each bound surfactant molecule, supporting the existence of positive $\mu_e$ in the studied region. It is worth mentioning that the slight increase of PDADMAC – SLES mixtures $\mu_e$ for the highest surfactant concentrations can be considered a signature of the appearance of surfactant micelles or hemicelles bridging several PDADMAC chains. This leads to the increase of the number of charged monomers in the formed complexes.

The above results are better explained on the basis of the binding of surfactant molecules to polymer monomers. The binding isotherm was obtained comparing the electromotive force, EMF, dependence on the SLES concentration for solutions of pure surfactant and mixtures according to the approach described by Mezei and Meszaros[56] (Figure 1b). Figure 1c shows that the surfactant binding to the polyelectrolyte chains is relatively high over a broad range of compositions, with the amount of surfactants remaining free in solution being lower than 1% for SLES concentrations below 0.2 mM. This could be considered, at least qualitatively, in agreement with the results reported by Campbell et al.[10, 27] for mixtures of PDADMAC and SDS. They found, at charge neutralization (~ 0.2 mM), a value of $\beta$ about 0.3 for a PDADMAC concentration of 100 ppm. The extrapolation of such results, assuming that $c_s^{free}$ does not change significantly, to a situation in which the concentration of polymer is 50 times higher, and similar to that here used, leads to a $\beta \sim 0.01$ in the isoelectric point. Hence for surfactant concentrations lower than those corresponding to the isoelectric point $\beta$ should assume a value well below 0.01, in agreement with our results. This gives an additional justification to $\mu_e$ values discussed above. Furthermore, the binding isotherm indicates that even for the highest SLES concentrations; the amount of surfactant molecules remaining free in solution is not larger than a 7 – 8 % of the total. Thus, it is possible to assume the existence of an almost negligible free surfactant concentration in solution, which makes possible to consider in the following the fulfilling of the zero free surfactant condition in analogy to that discussed by Llamas et al.[49] for PDADMAC – SLMT mixtures.

PDADMAC – SLES mixtures

Turbidity measurements (see Figure 2) provide useful insight for understanding the bulk aggregation. The absence of appreciable turbidity in samples with SLES concentrations below 0.2 mM allows one to consider those samples as belonging to an equilibrium one phase region. For higher surfactant concentrations, a sharp increase of the turbidity was found, which is related to the onset in the two phase region. It is worth mentioning that the increase of the turbidity cannot be ascribable to the formation of charge compensated complexes but it is probably related to the increase of their size by the formation of complexes involving several polymer chains.[4]

This seems to be in agreement with the sharp increase found in $R_H^{app}$ in the same concentration range (Figure 2). This scenario contrasts with most studies in literature, which suggested that the turbidity arises from the appearance of neutral complexes during the charge inversion transition from complexes in which their charge is governed by the polymer to other in which their charge is governed by the surfactant excess.[10, 20, 27] However, the obtained results indicate that the system is far from the neutralization region, thus the most likely explanation for the turbidity increase is the formation of kinetically-trapped aggregates during the preparation process of the samples.

PDADMAC – SLES mixtures

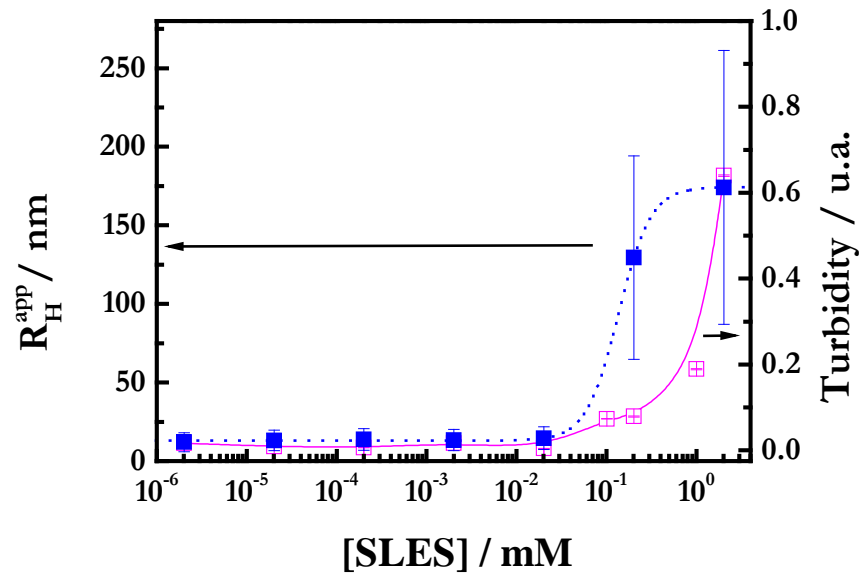

Figure 2. Dependences on SLES concentration of the turbidity (right axis, □) and of $R_H^{app}$ (left axis, ■) for PDADMAC – SLMT mixtures. Lines represent guides for the eyes.

### 3.2. Equilibrium adsorption at the water / vapor interface

Surface tension is a powerful tool for analyzing the adsorption of surface active compounds at fluid interfaces. PDADMAC solutions do not present any significant surface activity in the concentration studied here.[49] However, it is expected that both pure SLES solutions and PDADMAC – SLES mixtures present surfactant concentration dependent surface activity. Figure 3 shows the surface tension dependences on the SLES concentration obtained using different tensiometers for pure surfactant solutions and for mixtures of SLES and PDADMAC.

PDADMAC – SLES mixtures

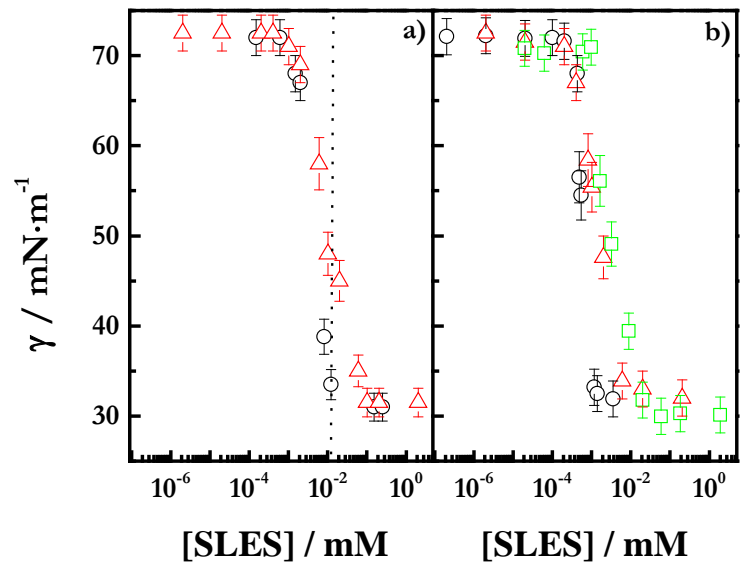

Figure 3. Surface tension dependences on SLES concentration for pure SLES solutions (a) and for PDADMAC – SLES mixtures (b) as were obtained using different tensiometers. Different symbols represents the results obtained using different techniques: Wilhelmy plate tensiometers with Pt probe (□) and with paper probe (○), and drop profile analysis tensiometer (Δ). The dotted line in panel a marks the cmc of pure SLES.

The dependences of γ on the SLES concentration obtained using different techniques present qualitative good agreement within the combined errors. For the lowest SLES concentrations, the behavior for pure surfactant solutions and for PDADMAC – SLES mixtures was similar, which can be explained assuming a low coverage of the fluid interface. Thus, the surface concentration is not high enough to reduce the interfacial free energy, and consequently γ remains close to the value expected for the bare water / vapor interface. The increase of the surfactant concentration leads to a monotonic decrease of γ for both pure surfactant solutions and for mixtures. This decrease starts at lower concentrations, almost one order of magnitude,

PDADMAC – SLES mixtures

for PDADMAC – SLES mixtures that for SLES solutions, which points out a synergetic effect on the decrease of the surface tension due to the interaction between the polymer and the surfactant. Similar behavior has been previously reported for the adsorption of other polyelectrolyte – surfactant mixtures at the water/vapor interface.[4, 10, 20, 27]

The detailed analysis of the surface tension isotherms obtained using different tensiometers provide important insights on the complex interfacial behavior of oppositely charge polyelectrolyte – surfactant solutions. In contrast to the behavior previously found for PDADMAC – SLMT solutions,[49] the results obtained using different tensiometers show good qualitative agreement within the combined error bars, and the existence of non-regular trends on the surface tensions were not found for PDADMAC - SLES mixtures, neither surface tension fluctuations[49] nor surface tension peaks.[27] Thus, it is expected that the kinetically trapped aggregates formed in the bulk can dissociate and spread at the interface.[14, 57] The adsorption of this kinetically trapped aggregates at the fluid interfaces can be explained assuming that the studied system is far from the precipitation region in the whole surfactant concentration range studied, and thus the aggregates formed in the bulk can remain trapped without any evidence of sedimentation.[26,28] This is supported by turbidity measurements shown in Figure 2. It is worth mentioning that the behavior found for PDADMAC – SLES mixtures is different to that of PDADMAC – SLMT one. In the latter mixtures, aggregates remain embedded at the interface, leading to surface tension fluctuactions.[49] These differences could be probably explained in terms of the differences on the molecular structure of SLES and SLMT. Thus, for SLMT molecules the hydrophobic tail is an alkyl chain which tends to minimize the number of contact points with water by the formation on compact aggregates that do not dissociate at the interface. On the other side, SLES contains oxyethylene groups within the hydrophobic chains. These groups can form hydrogen bonds with water, contributing to the dissociation of complexes once they are adsorbed at the interface.

In order to deepen on the physico-chemical aspects associated with the adsorption process of polymer – surfactant mixtures, Figure 4a shows the surface pressure, $\Pi(c) = \gamma_0 - \gamma(c)$, where $\gamma_0$ and $\gamma(c)$ represents the surface tension of the bare water / vapor interface and the surface tension of the solution / vapor interface, respectively, isotherms for SLES and PDADMAC – SLES mixtures obtained using a drop shape analysis tensiometer. It is worth recalling that results obtained using different tensiometers were in qualitative good agreement within the combined error bars. Thus, the conclusions obtained from the analysis of the drop profile tensiometer results can be extended to the measurements performed using other tensiometers.

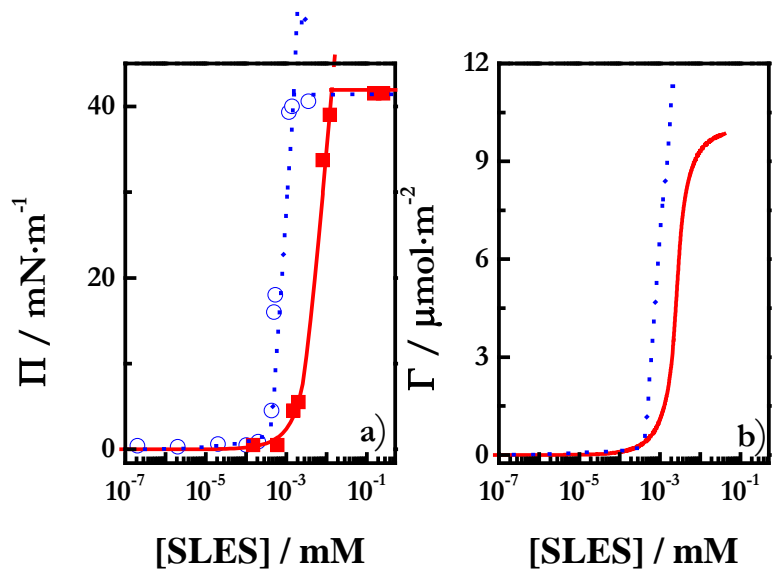

Figure 4. (a) Surface pressure isotherm for SLES (■,─) and PDADMAC – SLES (○, ····) obtained using a drop shape analysis tensiometer. Symbols represent experimental results and lines correspond to calculated isotherms. (b) Total surface excess ($\Gamma$) for SLES (─) and PDADMAC – SLES (····) layers calculated from the analysis of the surface pressure isotherms using the fits of the experimental results to the Frumkin isotherm for SLES PDADMAC – SLES mixtures

solutions and to the set of empirical equations described in the text for PDADMAC – SLES mixtures.

From the experimental isotherm, it is possible to estimate a critical micellar concentration (cmc) for SLES around $10^{-2}$ mM. The theoretical analysis of the experimental isotherm obtained for the adsorption of SLES solutions can be carried out in terms of the Frumkin isotherm (see Figure 4a),[58]

$$bc = \frac{\Gamma \omega}{1-\Gamma \omega} e^{-2a\Gamma \omega} \qquad (2)$$

$$-\frac{\Pi \omega}{RT} = \ln(1-\Gamma \omega) + a(\Gamma \omega)^2 \qquad (3)$$

where c and $\Gamma$ are the bulk concentration and the surface excess, respectively, $\omega$ corresponds to the area available per each surfactant molecule when the interface is saturated, b is referred to the adsorption equilibrium constant, *a* is the parameter of molecular interaction, and R and T are the gas constant and the absolute temperature, respectively. Figure 4a shows that the experimental isotherm for surfactant adsorption is well described in terms of a Frumkin isotherm with the parameters shown in Table 1.

Table 1. Parameters of the calculated isotherms describing the interfacial adsorption of SLES (Frumkin model) and mixtures of PDADMAC – SLES (empirical isotherm).

| **SLES** | **PDADMAC – SLES mixtures** |
|---|---|
| $10^{-5}\ \omega(m^2/mol) = 1$ | $10^{-5}\ \omega_{10}(m^2/mol) = 1.1$ |

PDADMAC – SLES mixtures

| | |
|---|---|
| $a = 1.48$ | $10^{-6}\,\omega_2\,(m^2/mol) = 9.5$ |
| $10^{-1}\,b\,(l/mmol) = 8.79$ | $10^{-3}\,b\,(l/mmol) = 1.668$ |
| | $\alpha = 0.8$ |
| | $10^3\,\varepsilon\,(m/mN) = 8.2$ |

For PDADMAC – SLES mixtures (see Figure 4a), a more complex scenario is expected, as was previously reported for PDADMAC - SLMT mixtures.[50] Therefore, the development of a theoretical model accounting for experimental results requires further considerations. Several theoretical models accounting for the adsorption of protein – surfactant mixtures at fluid interfaces can be found in the literature.[59-60] However, such models are not applicable to the mixtures studied here because of two main reasons. The first one is that PDADMAC, in comparison with most proteins, has a negligible surface activity, and the second one is the high polymer concentration.[61] Thus, some of the conditions implicit in the theoretical models are violated. Furthermore, it might be expected that the composition of the interface include polymer – surfactant complexes and free surfactant molecules.

The aforementioned issues make challenging the theoretical description of polyelectrolyte – surfactant mixtures and we have limited our description to the application of a set of empirical equations that provides a fitting of $\gamma_{eq}$ with a set of concentration-independent parameters. This methodology gives also information about the total surface excess and the dilational elasticity and has been used in a previous work.[50] It is worth mentioning that the above discussed aspects limit the application of any theoretical description based on a thermodynamics framework, thus the theoretical description is aimed to provide a qualitative

PDADMAC – SLES mixtures

description of the results in order to stress the complexity of the adsorption of polyelectrolyte – surfactant mixtures at the water/vapor interface. The set of empirical equations used on the description of the adsorption of PDADMAC – SLES mixtures can be summarized as follows

$$bc = \frac{\Gamma_2 \omega}{(1-\Gamma\omega)^{\omega_2/\omega}} \tag{4}$$

$$-\frac{\Pi\omega}{RT} = \ln(1-\Gamma\omega) \tag{5}$$

where $\Gamma_i$ (i = 1, 2) and $b = b_1$ are the surface excesses for molecules in different states and adsorption equilibrium constant in state 1, respectively. The total surface excess $\Gamma$ and the mean area per molecule $\omega$ are defined as follows

$$\Gamma = \Gamma_1 + \Gamma_2 \tag{6}$$

$$\omega\Gamma = \theta = \omega_1\Gamma_1 + \omega_2\Gamma_2 \tag{7}$$

where $\theta$ is the surface coverage. From the set of empirical equations it is possible to calculate the ratio of adsorbed molecules in the different states as follows

$$\frac{\Gamma_2}{\Gamma_1} = \exp\left(\frac{\omega_1 - \omega_2}{\omega}\right)\left(\frac{\omega_1}{\omega_2}\right)^\alpha \exp\left(-\frac{\Pi(\omega_1 - \omega_2)}{RT}\right) \tag{8}$$

where $\alpha$ is a constant accounting for different on the adsorption in each states. A last aspect to consider is the role of the interfacial compressibility $\varepsilon$ of the molecules in state 1,

$$\omega = \omega_0\left(1 - \varepsilon\Pi\theta\right) \tag{9}$$

where $\omega_0$ is the area per molecule at $\Pi = 0$. It is worth mentioning that the parameters (see Table 1) obtained applying this set of equations to the adsorption of PDADMAC – SLES PDADMAC – SLES mixtures mixtures are far to provide a real physical picture of the interfacial behavior of the studied systems. However, it allows us to stress the complexity of the interfacial behavior of polyelectrolyte – surfactant mixtures. Figure 4a shows that the curve calculated for PDADMAC – SLES on the bases of the set of empirical equations provides a good description of the experimental isotherm, despite the complexity of the interfacial behavior.

Figure 4b shows the total surface excesses calculated from the analysis of the surface isotherm on the bases of the above equations. In both SLES and PDADMAC – SLES mixtures, the calculated surface excesses present a monotonic increase with the surfactant concentration. This is the typical behavior expected for the adsorption of pure surfactant. For PDADMAC – SLES mixtures, the dependence shows a qualitative good agreement with that found in PDADMAC – SLMT mixtures.[50] However, the accurate description of the polyelectrolyte – surfactant mixtures systems requires the development of a complex thermodynamic multi-component model accounting for the differences on the activities of the different components.

### 3.3. Adsorption kinetics at the water / vapor interface

Better understanding of the adsorption mechanism of PDADMAC – SLES can be obtained from the analysis of the adsorption kinetics at the water/vapor interface. Information about this aspect can be obtained from the analysis of the time dependences of the surface tension (dynamics surface tensions) measured using the drop shape analysis tensiometer. Figure 5 shows the dynamics surface tension for the adsorption of SLES (Figure 5a) and PDADMAC – SLES mixtures (Figure 5b) at the water / vapor interface for solutions containing different surfactant concentrations.

PDADMAC – SLES mixtures

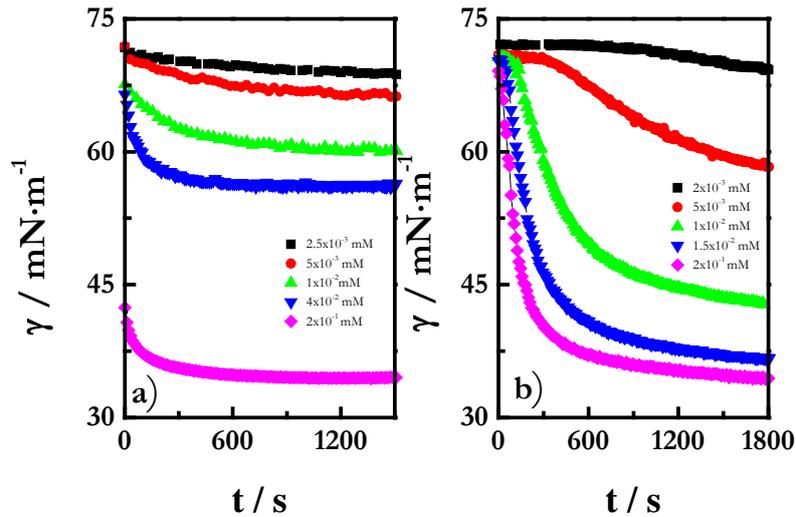

Figure 5. (a) Dynamic surface tension for SLES solutions with different concentrations. (b) Dynamic surface tension for PDADMAC - SLES solutions with different concentrations.

The results show a qualitatively similar trend for the adsorption of both polyelectrolyte - surfactant mixtures and pure surfactant solutions, with equilibration times decreasing as the surfactant concentration increases. However, the equilibration of PDADMAC – SLES layers requires, independently of the SLES concentration, longer times than the adsorption of pure SLES. Furthermore, it is worth mentioning that whereas the adsorption of SLES at the water/vapor interface presents a monotonous decay with time over the whole concentration range, an induction time was found for the adsorption of PDADMAC – SLES mixtures. After such induction time, the surface tension starts to decay a monotonous with time. The induction region is reduced as the surfactant bulk concentration increases in similar way to that described by Erickson et al.[62]

The induction time can be explained, similarly to that occurring in the adsorption of protein at fluid interfaces,[62] as a consequence of a conformational transition of the complex once they

PDADMAC – SLES mixtures

are attached at the fluid interface. Thus, polymer – surfactant complexes adsorb at the interface as kinetically trapped aggregates which dissociate and spread, leading to the surface tension drop.[14, 57]

The decrease of the induction time with the increase of the SLES concentration can be explained assuming an enhanced adsorption of the kinetically trapped aggregates. This leads to a fast crowding of the interface, shortening the time necessary to overcome the surface excess threshold of kinetically trapped aggregates which lead to a prior surface tension drop. Therefore it is possible to assume that the induction time is related to the time necessary for overcoming a threshold surface excess value of kinetically trapped aggregates at the interface. This scenario contrasts with the fact that in the adsorption at the water/vapor interfaces of PDADMAC-SLMT mixtures in which dissociation and spreading were not found, and the kinetically trapped aggregates remain compact at the interface as shown by Brewster Angle Microscopy images.[49] The different mechanisms involved in the equilibration of PDADMAC – SLES and PDADMAC – SLMT layers provide an explanation for the different time scales involved in their adsorption. Whereas in PDADMAC – SLMT mixtures the surface tension drops due to the incorporation of isolated complexes which coalesce as the interfacial concentration increases, for PDADMAC – SLES mixtures the surface tension decrease is associated with the dissociation and spreading of the adsorbed complexes. Therefore, the surface tension decrease for PDADMAC – SLMT is mainly governed by the diffusion of the colloidal aggregates from the bulk to the interface whereas for PDADMAC – SLES is controlled by the complex dissociation and spreading process which is expected to involve shorter time scales.

Another difference between the results obtained for PDADMAC – SLES and PDADMAC – SLMT is that the appearance of the induction time does not allow us to provide a theoretical description of the adsorption kinetics of the complexes by the combination the Ward-Tordai PDADMAC – SLES mixtures equation[63] ($\Gamma(t) = 2\sqrt{\dfrac{D}{\pi}}\left[c_0\sqrt{t} - \int_0^{\sqrt{t}} c\left(0, t-t'\right)d\sqrt{t'}\right]$). In this equation D is the diffusion coefficient, $c_0$ represents the initial bulk concentration, t corresponds to the time and t' is a dummy integration variable) and the set of empirical equation describing the surface tension isotherm.

### 3.4. Dilational rheology

A last point with interest for understanding the physico-chemical properties of systems under technologically relevant conditions is the study of its response against an external mechanical perturbation.[34] This provides information about the relaxation processes appearing at the interface which play an important role in the development of optimized industrial application, e.g. thin film deposition or stabilization of dispersed systems.[16, 64] In the following, we will discuss the dependences of the dilational viscoelastic moduli $\varepsilon^* = \varepsilon' + i\varepsilon''$ (ε′ represents the dilational elastic modulus and ε″ the viscous modulus) on the surfactant concentration and the deformation frequency.[65] This is essential for a better understanding of the complex mechanisms involved in the equilibration of the interface. Furthermore, it can help us to provide a more detailed description of the physico-chemical bases underlying the adsorption processes of polymer – surfactant at fluid interfaces. It is worth mentioning that both PDADMAC – SLES mixtures and SLES solutions present ε′ values more than one of magnitude larger than those of ε″. Thus, for the sake of simplicity we will only discuss the dependences of the elastic modulus.

The study of the dilational rheology of interfacial layers is important to understand the processes leading to the equilibration of the interface. Furthermore, this type of studies can help to understand the complex physico-chemical framework associated with the adsorption processes of polyelectrolyte – surfactant mixtures at the fluid interface. Figure 6 shows the PDADMAC – SLES mixtures frequency dependences of the elastic modulus for solutions of SLES and PDADMAC – SLES as were obtained by the oscillatory method barrier.

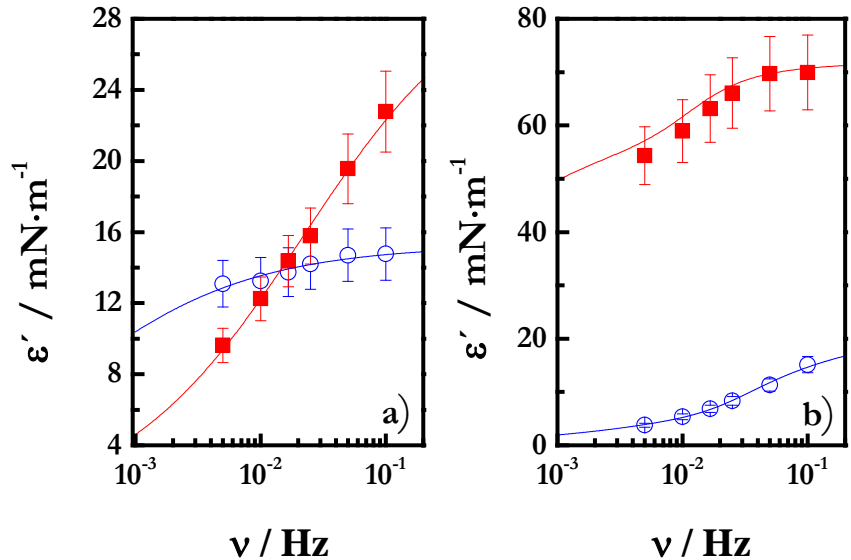

Figure 6. Frequency dependences of the elastic modulus for SLES solutions (a) and PDADMAC – SLES mixtures (b) obtained for two different SLES concentrations. In the panels, the symbols are referred to different SLES concentrations (squares correspond to 0.002 mM and circles correspond to 0.1 mM), whereas lines represent the best fit of the experimental curves to theoretical models.

The results evidence that for the lowest surfactant concentrations, $\varepsilon'$(SLES) ~ $\varepsilon'$(PDADMAC – SLES mixtures), whereas the opposite is true for the highest surfactant concentrations $\varepsilon'$( PDADMAC – SLES mixtures) > $\varepsilon'$( SLES). The frequency dependence of the response upon dilation for both SLES solutions and PDADMAC – SLES mixtures is that typically expected for adsorption layers of surface active material at fluid interfaces.[32, 34] However, the detailed analysis of such dependences on terms of theoretical models evidences differences between PDADMAC – SLES mixtures the adsorption of surfactant solutions and polyelectrolyte – surfactant mixtures. The dynamic behavior of surfactant layers (Figure 6a) at the water/vapor interface can be explained on the bases of a diffusion controlled adsorption which can be described in terms of Lucassen – van den Tempel model.[66-67] This model describes the frequency dependence of the viscoelasticity modulus as follows

$$\varepsilon = \frac{1+\xi+i\xi}{1+2\xi+2\xi^2}\varepsilon_0 \qquad (10)$$

where $\xi = \sqrt{\frac{\nu_D}{\nu}}$ with $\nu_D$ being the frequency of the diffusion exchange and $\nu$ is the frequency of the deformation, and $\varepsilon_0$ is the Gibbs elasticity. For the case of PDADMAC – SLES mixtures, the scenario is more complex (Figure 6b), and the use of a diffusion-controlled mechanism is not able to account for the dynamics of the interfacial layers. This makes necessary to introduce an additional relaxation process on the description of the viscoelastic modulus in agreement with Ravera et al.[68-69] Thus, the dependence of the viscoelastic modulus on the frequency reads as follows

$$\varepsilon = \frac{1+\xi+i\xi}{1+2\xi+2\xi^2}\left[\varepsilon_0 + (\varepsilon_1 - \varepsilon_0)\frac{1+i\lambda}{1+\lambda^2}\right] \qquad (11)$$

where $\lambda = \nu_1/\nu$ with $\nu_1$ being the characteristic frequency of the additional relaxation process, and $\varepsilon_1$ the high frequency elasticity. The fact that the model described in Equation (11) can provide a description of the response against dilation deformation of PDADMAC – SLES layers may be easily correlated to the complexity of the adsorption process of such system which was previously discussed on the bases of the surface tension isotherm and the adsorption kinetics. Thus, the combination of the results discussed above with the new insights provide by the experiments on dilational rheology allows us to deepen of the PDADMAC – SLES mixtures mechanistic aspects involved in the equilibration of the interfacial layers. This is a further confirmation of the existence of adsorption mechanism involving two-steps. The first step corresponds to a diffusion-controlled adsorption of the kinetically trapped aggregates formed in the bulk at the subsurface, and the second one is governed by the Marangoni gradient associated with the reorganization, probably through dissociation and spreading, of the complexes adsorbed during the first step. The two-step mechanism for the adsorption of PDADMAC – SLES mixtures at the water/vapor interface agrees qualitatively with the description provided by Noskov et al.[41] of the adsorption process of PDADMAC–SDS mixtures. Figure 7 shows the concentration dependences of $\varepsilon_0$, $\varepsilon_1$, $\nu_D$ and $\nu_1$ obtained for interfacial layers of PDADMAC – SLES mixtures.

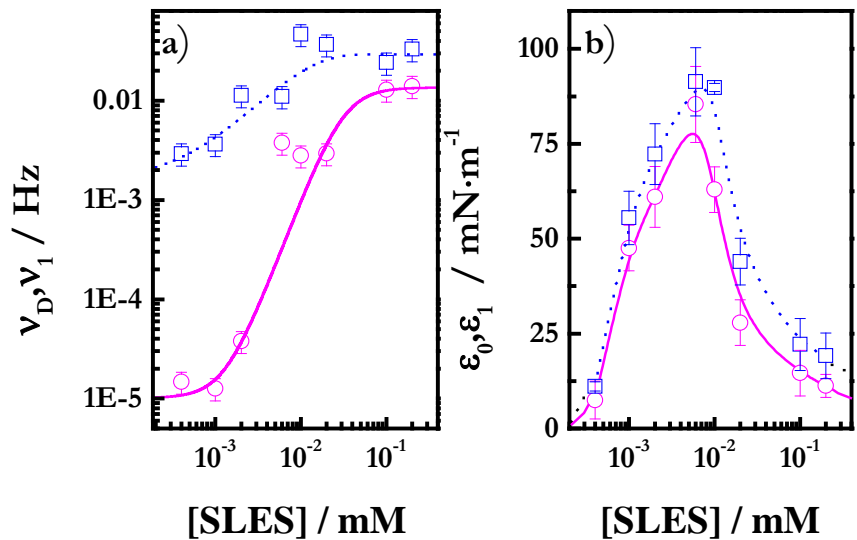

Figure 7. SLES concentration dependences for PDADMAC – SLES mixtures. (a) Frequencies of the different relaxation processes: $\nu_D$ (○, —) and $\nu_1$ (□, ⋯). (b) Limit elasticities: $\varepsilon_0$ (○, —) and $\varepsilon_1$ (□, ⋯). For each of the two panels, the symbols represent the experimental data and the lines are guides for the eyes.

PDADMAC – SLES mixtures

The results show that within the SLES concentration range considered, the frequency associated with the interfacial relaxation, $\nu_1$, presents values higher than that corresponding to the diffusion-controlled transport, $\nu_D$. This may be easily understood assuming that the reorganization associated with $\nu_1$ occurs only when the material is adsorbed at the interface. In addition, both $\nu_D$ and $\nu_1$ increases with the surfactant concentration. This may be explained in the case of $\nu_D$ considering the enhanced surface activity of the formed complex with the surfactant concentration increases. The dependence of $\nu_1$ is explained considering that the increase of the surfactant concentration in the bulk increases the surface excess of complexes at the interface. These complexes act as reservoirs of material at the interface that can dissociate and spread at the interface, with this process being favored with the increase of the surfactant concentration.[70] The analysis of the limit elasticities (Figure 7b) shows that $\varepsilon_1 > \varepsilon_0$ over the whole SLES concentration range. In addition, the analysis of the concentration dependences of $\varepsilon_0$ and $\varepsilon_1$ increase from values close to zero to reach a maximum, and then the limit elasticities drop again to values around zero. This is the behavior typically expected for the surfactant concentration dependence of the elastic modulus.[42] Similar concentration dependences are found for the elastic modulus obtained from experiments performed at different deformation frequency. For the sake of example, Figure 8 shows the concentration dependences of the elastic moduli obtained at different deformation frequencies. It is worth mentioning that other frequencies have similar dependences.

PDADMAC – SLES mixtures

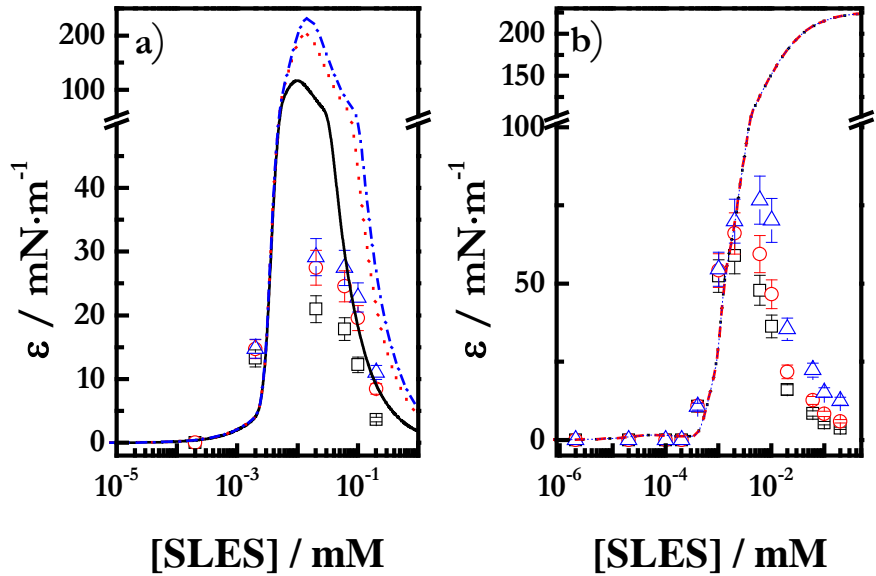

Figure 8. (a) Concentration dependences of the elastic modulus for SLES adsorption layers as were obtained from oscillatory barrier experiments performed at different frequencies. (b) Concentration dependences of the elastic modulus for PDADMAC - SLES adsorption layers as were obtained from oscillatory barrier experiments performed at different frequencies. In both panels, the symbols represent the experimental results and the lines represent the theoretical curves for different frequencies: □ and ─ for ν = 0.01 Hz, ○ and ··· for ν = 0.05 Hz, △ and ─·─ for ν = 0.1 Hz.

The theoretical curves for SLES layers calculated on the bases of the parameters obtained using the Frumkin isotherm (see Table 1) provides a qualitatively description of the dependence of the elastic modulus on the surfactant concentration. However, the values calculated overestimate the experimental ones, describing the low frequency region, and when the surfactant concentration gets closer to the cmc of the surfactant. In the case of PDADMAC – SLES, the calculated elasticity modulus does not describe the experimental

PDADMAC – SLES mixtures

data except for the lowest SLES concentrations. This again points out the high complexity of the considered mixture, which requires a more detailed theoretical modelling.

## 4. Conclusions

This work, by mean of equilibrium and dynamics surface tension measurements, has shown the complex physico-chemical bases associated with the adsorption of mixtures formed by a polycation PDADMAC and the oppositely charged surface surfactant SLES at the water / vapor interface. The results have pointed out that the interaction between polyelectrolytes and oppositely charged surfactants increase the complexity of the adsorption phenomena in relation to that of the pure surfactant.

The results have shown that the mechanism of adsorption of PDADMAC – SLES mixtures at the water/vapor interface can be interpreted to take place in two steps. The first one is associated with the diffusional transport of the kinetically trapped aggregates from the bulk to the vicinity of the interface, and the second one to an interfacial reorganization associated with the complex dissociation and spreading of the aggregates adsorbed at the interface. The complexity of this behaviour has been described qualitatively using a set of empirical equations, which accounts for the existence of two relaxation process in dynamics measurements. However, this description does not provide a satisfactory explanation of the experimental results which underline the complexity of the mechanism involved in the adsorption of polyelectrolyte – surfactant mixtures at the liquid/vapor interface. This makes necessary to address the theoretical modelling of these systems, including their specific characteristics order to provide a comprehensive explanation of the most fundamental bases of their behavior to optimize the design and

PDADMAC – SLES mixtures

fabrication of new polyelectrolyte – surfactant mixtures with technological interest.

**Conflict of interest**

The authors declare no competing financial interest.

**Acknowledgements**

This work was funded in part by MINECO under grants CTQ-2016- 78895-R and by the EU under ITN Marie Curie - CoWet (Grant Agreement 607861). The authors are grateful to CAI of Spectroscopy (Universidad Complutense de Madrid) for the use of their facilities.

PDADMAC – SLES mixtures

PDADMAC – SLES mixtures

PDADMAC – SLES mixtures

PDADMAC – SLES mixtures

PDADMAC – SLES mixtures

PDADMAC – SLES mixtures

**TOC Graphics (For Table of Content Use Only)**

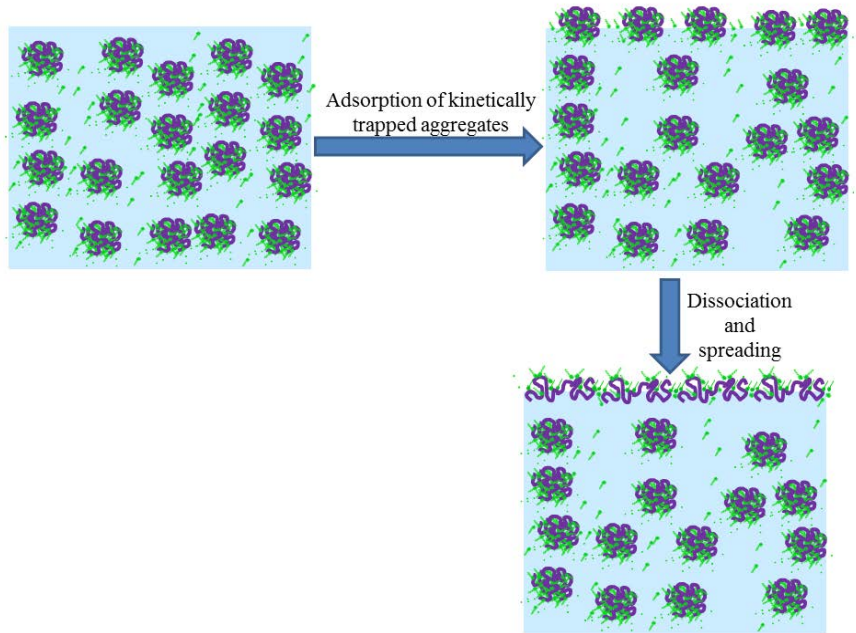

PDADMAC – SLES mixtures